\documentclass{aa}
\usepackage{graphicx}
\usepackage{natbib}
\newcommand{\Omegav}{\mbox{\boldmath$\Omega$}}
\newcommand{\xiv}{\mbox{\boldmath$\xi$}}
\begin{document}

\title{Hall drift of axisymmetric magnetic fields in solid neutron-star matter}

   \author{Andreas Reisenegger\inst{1,2}
          \and
          Rafael Benguria\inst{3}
          \and
          Joaqu{\'\i}n P. Prieto\inst{1}
          \and
          Pablo A. Araya\inst{1}\thanks{Current address: Kapteyn Astronomical Institute, University of Groningen, P. O. Box 800, 9700 AV, Groningen, The Netherlands}
          \and
          Dong Lai\inst{4}
          }

   \offprints{A. Reisenegger}

   \institute{Departamento de Astronom{\'\i}a y Astrof{\'\i}sica,
             Pontificia Universidad Cat\'olica de Chile, Casilla 306, Santiago 22,
             Chile\thanks{Permanent address of A.~Reisenegger} \email{areisene@astro.puc.cl}\\
         \and
             Max-Planck-Institut f\"ur Astrophysik, Karl-Schwarzschild-Str. 1,
             85741 Garching bei M\"unchen, Germany\\
         \and
             Departamento de F{\'\i}sica,
             Pontificia Universidad Cat\'olica de Chile, Casilla 306, Santiago 22,
             Chile\\
         \and
             Center for Radiophysics and Space Research, Department of Astronomy,
             Cornell University, Ithaca, NY 14853, USA\\
             }

   \date{Received ; accepted }

\abstract {The Hall drift, namely, the transport of magnetic flux
by the moving electrons giving rise to the electrical current, may
be the dominant effect causing the evolution of the magnetic field
in the solid crust of neutron stars. It is a nonlinear process
that, despite a number of theoretical efforts, is still not fully
understood.} {Through mostly analytic arguments and solutions, we
intend to help understand this highly nonlinear process.} {We use
the Hall induction equation in axial symmetry to obtain some
general properties of nonevolving fields, as well as analyzing the
evolution of purely toroidal fields, their poloidal perturbations,
and current-free, purely poloidal fields. We also analyze energy
conservation in Hall instabilities and write down a variational
principle for Hall equilibria.}{We show that the evolution of any
toroidal magnetic field can be described by Burgers' equation, as
previously found by Vainshtein and collaborators in a
plane-parallel geometry. This evolution leads to sharp current
sheets, which dissipate on the Hall time scale, yielding a
stationary field configuration that depends on a single, suitably
defined coordinate. This field, however, is unstable to poloidal
perturbations, which grow as their field lines are stretched by
the background electron flow, as in the instabilities found
numerically by Rheinhardt and Geppert. On the other hand,
current-free poloidal configurations are stable and could
represent a long-lived crustal field supported by currents in the
fluid stellar core. There may be additional, stable
configurations, corresponding to restricted local minima or maxima
of the magnetic energy.}{Hall equilibria can be described by a
simple variational principle. Long-lived, toroidal fields are not
expected in neutron star crusts or other regions where Hall drift
is the dominant evolution mechanism. However, other stable
configurations do exist, such as current-free poloidal fields and
possibly others.}


\titlerunning{Hall drift}
\authorrunning{A. Reisenegger et al.}

   \maketitle
%

\section{Introduction}
\label{sec:intro}

The Hall drift, namely the advection of magnetic flux by the
current associated with it, is important in systems where the
magnetic field is strong enough to make the cyclotron frequency
comparable or greater than the collision rate, and bulk flow
velocities are not much larger than the relative velocity of
different charge carriers, associated to the electric current. For
some time, it has been realized that this effect might play an
important role in the evolution of neutron star magnetic fields,
which under some circumstances are large enough for the above
conditions to be satisfied, particularly in the solid crust of
these stars, in which the electrons are the only mobile charges
\citep{Jones,GR92,Cumming}.

The evolution of the magnetic field under the Hall effect is a
nonlinear process that has so far eluded a full theoretical
understanding. Since the Hall effect conserves the magnetic energy
\citep{Urpin,GR92}, it has been argued that it will act mainly
through the generation of steep magnetic field gradients, on which
resistive dissipation can act much more quickly than it would on a
smooth field. \citet{GR92} gave general arguments showing that
this could happen through a turbulent cascade transferring energy
from larger to smaller scales, which was later supported by
simulations \citep{Biskamp}. On the other hand, \citet{VCO}
considered an analytic model problem of a plane-parallel slab of
matter with a vertical electron density gradient, showing that the
evolution of a purely horizontal, everywhere parallel magnetic
field can be described by Burgers' equation, producing
discontinuities (current sheets) that are smoothed if dissipation
is included, but indeed lead to rapid dissipation of the magnetic
field.

Other attempts at a better understanding have been numerical. Some
authors \citep{Naito,Hollerbach02,Hollerbach04} have computed the
evolution of the first few eigenmodes of the magnetic diffusion
equation in spherical symmetry, nonlinearly coupled to each other
by the Hall effect. \citet{Urpin} have modelled the evolution of
the magnetic field by solving the evolution equation on a grid for
a purely toroidal field on a uniform density sphere, again finding
that the magnetic field develops sharper gradients, which later
dissipate by resistive effects. Simulating the evolution of a
poloidal field in a similar way \citep{Shalybkov}, energy was
found to be transferred to a toroidal component and back,
complicating the results substantially. \citet{Pons} have recently
simulated the evolution of both poloidal and toroidal components
in a neutron star crust with a code that combines finite
differences in the radial direction with a spherical harmonic
decomposition in angle, using a realistic description of the
radial profile of electron density and electric resistivity and
the evolution of the latter. They confirm the transfer of energy
back and forth between the poloidal and the toroidal component
(especially from the former to the latter), and find regimes along
the neutron star evolution when Hall drift or Ohmic diffusion
dominate.

In a different approach, \citet{Rheinhardt02} (see also
\citealt{Geppert02,Geppert03,Rheinhardt04}) did a numerical
stability analysis on plane-parallel field configurations that do
not by themselves evolve under the Hall effect, finding both
stable modes (a particular case of which are the well-known
``whistler waves'') and unstable ones. Based on several examples,
they conjecture that, at least in simple geometries, the
determining factor for the existence of unstable modes is a
sufficiently large second spatial derivative of the unperturbed
magnetic field strength.

The present paper reports an analytical study of the evolution of
a magnetic field with a possibly realistic, axially symmetric
configuration in a solid star with non-uniform electron density.
We first give a short re-derivation of the relevant form of the
induction equation (\S~\ref{sec:induction}) and specialize it to
the two independent scalar functions determining the magnetic
field in axial symmetry, discussing some of its implications
(\S~\ref{sec:axial}). Then, we consider the very special case of a
purely toroidal magnetic field (\S~\ref{sec:toroidal}). It is
shown that, if only the Hall drift is active, the magnetic field
evolves independently on each of a family of nested toroidal
surfaces, again being determined (for a suitable change of
variables) by Burgers' equation (\S~\ref{sec:Burgers}). As found
by \citet{VCO}, this leads to the formation of current sheets that
can rapidly dissipate, so the field evolves on a Hall time scale
to a simple configuration on each of these surfaces, minimizing
the magnetic energy while conserving the flux within each toroid
(\S~\ref{sec:conserved}). Next (\S~\ref{sec:instability}), we
consider various aspects of small perturbations and Hall
instabilities: We show that a stationary toroidal magnetic field
is generally unstable to poloidal perturbations
(\S~\ref{sec:poloidal} --- up to here, our results were already
summarized in \citealt{R05}) and interpret this through the
plane-parallel model considered by previous authors
(\S~\ref{sec:plane}), we discuss the issue of energy conservation
in Hall instabilities, show that current-free configurations are
stable (\S~\ref{sec:energy}), and that a current-free, poloidal
field supports whistler-like waves (\S~\ref{sec:current-free}).
Then, we show that Hall equilibria can be derived from a
variational principle, by requiring the magnetic energy to be
stationary with respect to small displacements that do not alter
the electron density, and use this to argue that some
configurations, corresponding to maxima or minima of the magnetic
energy subject to these constraints, must be stable with respect
to small perturbations evolving solely through the Hall drift
(\S~\ref{sec:variational}). Finally, we list our main conclusions
(\S~\ref{sec:conclusions}).


\section{Induction equation}
\label{sec:induction}

In order to clarify our assumptions and notation, we start by
re-deriving the equation of motion for the magnetic field in an
arbitrary geometry. More general derivations have been given,
among others, in \citet{GR92} and \citet{R05}.

We consider a material in which the electrons are the only moving
particles, embedded in a perfectly rigid, neutralizing background
(i.~e., an idealized crystal lattice) against which they can
occasionally scatter. The steady-state, local average velocity of
the electrons is determined by the balance of the Lorentz force
against the time-averaged momentum loss through collisions, which
yields a generalized Ohm's law,
\begin{equation}
\mathbf{j}=-ne\mathbf{v}=\sigma\left(\mathbf{E}+\frac{\mathbf{v}\times\mathbf{B}}{c}\right),
\end{equation}
where $\mathbf{j}$ is the electric current density, $n$, $-e$, and
$\mathbf{v}$ are the number density, charge, and average velocity of
the electrons, $\sigma$ is the (scalar) electrical conductivity,
$\mathbf{E}$ and $\mathbf{B}$ are the electric and magnetic field,
and $c$ is the speed of light. The evolution of the magnetic field
with time $t$ is described by the induction equation, $\partial
\mathbf{B}/\partial t=-c\nabla\times \mathbf{E}$, which, replacing
$\mathbf{E}$ from eq. (1), takes the form
\begin{equation}
\frac{\partial \mathbf{B}}{\partial
t}=\nabla\times\left(\mathbf{v}\times\mathbf{B}-\frac{c\mathbf{j}}{\sigma}\right).
\end{equation}
In the infinite-conductivity limit, this equation asserts that the
magnetic field lines can be thought of as drifting along with the
electron flow (``Hall drift''). The term involving the conductivity
corresponds to deviations from this idealization due to resistive
diffusion of the magnetic field.

In the slow-motion limit (i.~e., ignoring electromagnetic
radiation), the current, and thus the electron velocity, are also
related to the magnetic field by Amp\`ere's law,
$\mathbf{j}=-ne\mathbf{v}=c\nabla\times\mathbf{B}/(4\pi)$. This
yields a magnetic field evolution law with $\mathbf{B}$ as the only
dynamical variable,
\begin{equation}\label{induction}
\frac{\partial \mathbf{B}}{\partial
t}=-\nabla\times\left[\frac{c}{4\pi n
e}\left(\nabla\times\mathbf{B}\right)\times\mathbf{B}+\eta\nabla\times\mathbf{B}\right],
\end{equation}
where we have introduced the magnetic diffusivity, $\eta\equiv
c^{2}/(4\pi\sigma)$. The Hall term is quadratic in $\mathbf{B}$,
which has so far impeded a full theoretical understanding.

\section{Axial symmetry}
\label{sec:axial}

We now specialize to the case of an axisymmetric star, described
by the standard cylindrical coordinates $R$, $\phi$, and $z$, with
electron density $n(R,z)$. For simplicity, we set $\eta=0$ in this
derivation, although below we consider the effects of resistive
diffusion. The most general, axisymmetric magnetic field can be
decomposed into a toroidal component
\begin{equation}\label{toroidal}
{\bf B}_T={\cal B}(R,z)\nabla\phi
\end{equation}
and a poloidal component
\begin{equation}\label{poloidal}
{\bf B}_P=\nabla{\cal A}(R,z)\times\nabla\phi.
\end{equation}
This decomposition makes it explicit that the field depends only
on two scalar functions, ${\cal B}$ and ${\cal A}$, and explicitly
satisfies the condition of zero divergence independently for both
components. We choose to write it in terms of the gradient
$\nabla\phi=\hat\phi/R$ instead of the unit vector $\hat\phi$ in
order to make easy use of the identity $\nabla\times\nabla\phi=0$.
For future reference, we also write the toroidal and poloidal
components of the electron velocity,
\begin{eqnarray}\label{velocity}
{\bf v}_T=&{c\over 4\pi en(R,z)}\Delta^*{\cal A}~\nabla\phi,\\
{\bf v}_P=&-{c\over 4\pi en(R,z)}\nabla{\cal B}\times\nabla\phi,
\end{eqnarray}
where $\Delta^*\equiv\nabla^2-(2/R)\partial/\partial R$ is the
Grad-Shafranov operator (e.~g., \citealt{Kulsrud}, \S~4.9).

This shows that the magnetic field lines lie on the surfaces
${\cal A}=\mathrm{constant}$, while the current or electron flow
lines lie on surfaces ${\cal B}=\mathrm{constant}$. If both
$\mathcal{A}$ and $\mathcal{B}$ are taken to be zero on the
symmetry axis, then $2\pi{\cal A}$  is the poloidal flux enclosed
by a given surface ${\cal A}=\mathrm{constant}$, whereas $c{\cal
B}/2$ is the total current enclosed by the corresponding surface
(see also \citealt{Kulsrud}, \S~4.9).

Replacing the previous expressions into the induction equation,
eq.~(\ref{induction}), and separating toroidal and poloidal
components, we obtain the evolution of the two scalar functions,
\begin{eqnarray}\label{evolution}
{\partial{\cal B}\over\partial
t}&=&R^2\left[\nabla(\chi\Delta^*{\cal A})\times\nabla{\cal
A}+\nabla\chi\times{\cal B}\nabla{\cal B}\right]\cdot\nabla\phi,\\
{\partial{\cal A}\over\partial t}&=&R^2\chi\nabla{\cal
A}\times\nabla{\cal B}\cdot\nabla\phi,
\end{eqnarray}
where we have introduced the scalar function $\chi(R,z)\equiv
c/[4\pi e n(R,z)R^{2}]$. These equations show that the toroidal
and poloidal components of the field couple strongly to each other
in their evolution, so the latter might become quite intricate in
the general case. However, there are a few simple but interesting
things to be noted.

First, ${\cal A}$ does not evolve if the surfaces ${\cal
A}=\mathrm{constant}$ and ${\cal B}=\mathrm{constant}$ coincide,
i.~e., one of these scalar functions can be written as a function
of the other. In this case, the poloidal components of the
magnetic field and the current density are parallel to each other
at every point, ${\bf j}_P=(c/4\pi)(d\mathcal{B}/d\mathcal{A}){\bf
B}_P$. They lie in these surfaces, and the coefficient is also
constant on each surface. Since there cannot be substantial
currents flowing into and out of the star, the poloidal currents
on any field line extending out of the star must vanish
identically.

This condition, ${\bf j}_P\times{\bf B}_P=0$, is equivalent to
imposing that the azimuthal component of the Lorentz force is
zero. Since neutron stars are born in a fluid state, it might be
natural to require that they start in an MHD equilibrium, in which
magnetic forces are balanced by fluid forces. In an axially
symmetric configuration, the fluid forces cannot have an azimuthal
component, therefore naturally leading to this condition.

Second, the previous condition does not ensure that ${\cal B}$
will not evolve, so it may not continue to be satisfied at later
times as the magnetic field Hall-drifts through the solid (see
\citealt{RT07} for a related discussion for Hall drift in a fluid
star). In order to insure that ${\cal B}=\mathrm{constant}$ as
well, one also needs to impose that $\chi(\Delta^*{\cal A}+{\cal
B}d{\cal B}/d{\cal A})$ is constant on the same surfaces as ${\cal
A}$ and ${\cal B}$. A particular case of a stationary solution is
the poloidal field configuration found by \citet{Cumming}, in
which the electron velocity corresponds to a pure rigid-body
rotation, so one has ${\cal B}=0$ and an angular velocity
$\Omega=\chi\Delta^*{\cal A}=\mathrm{constant}$ everywhere.

Aside from this very special case, a purely poloidal initial field
(${\cal A}\neq 0, {\cal B}=0$) will generate a toroidal component
as well, because its electron velocity corresponds to a
differential rotation that stretches the field lines in the
azimuthal direction. However, the converse is not true: If the
initial field is purely toroidal, i.~e., initially ${\cal A}$ is
exactly zero, it will remain that way, and only $\mathcal{B}$ will
evolve \citep{Urpin}. The evolution of such a field is much
simpler than the general case, but still non-trivial, so we devote
the next section to its study.


\section{Toroidal field}
\label{sec:toroidal}


\subsection{Evolution: Burgers equation and current sheets}
\label{sec:Burgers}

The evolution of a purely toroidal magnetic field, now again
including a resistive term with axisymmetric diffusivity,
$\eta(R,z)$, is given by
\begin{equation}\label{toroidalinduction}
\frac{\partial\mathcal{B}}{\partial t}+\mathbf{w}\cdot\nabla\mathcal{B}=R^{2}\nabla\cdot\left(\frac{\eta}{R^{2}}\nabla\mathcal{B}\right),
\end{equation}
with\footnote{There is an ambiguity in the choice of the vector
field $\mathbf{w}$, in the sense that adding to it another vector
field that is everywhere perpendicular to $\nabla\mathcal{B}$ does
not change the evolution of $\mathcal{B}$. Our choice is
convenient in the sense that the flow lines of $\mathbf{w}$ are
independent of $\mathcal{B}$ and therefore do not change with
time.} $\mathbf{w}\equiv
R^{2}\mathcal{B}\nabla\chi\times\nabla\phi$. Thus, in the absence
of dissipation ($\eta=0$), the quantity $\mathcal{B}$ can be
viewed as being advected by the ``velocity field'' $\mathbf{w}$
like a scalar conserved quantity by a hydrodynamic flow. This
advection is very different from that of the magnetic flux
$\mathbf{B}$ by the electron velocity field
$\mathbf{v}=R^{2}\chi\nabla\phi\times\nabla\mathcal{B}$, which can
change the magnitude of the field by compressing or diluting the
flux. The velocity field $\mathbf{w}$ is clearly perpendicular to
$\mathbf{B}$ (i.~e., poloidal and tangent to surfaces of
$\chi(R,z)=\mathrm{constant}$).

Figure~1 shows the surfaces of constant $\chi$ for a particular
neutron star model. The singular surface $\chi=+\infty$ is the
union of the star's symmetry axis (on which $R=0$) and its surface
(where $n=0$). For all $\chi$ larger than its minimum value
$\chi_{min}$, the ``$\chi$-surfaces'' are nested surfaces of
toroidal topology, while $\chi_{min}$ defines an equatorial
circle.
\begin{figure}[!ht]
\centering
\includegraphics[width=8cm,height=6.5cm]{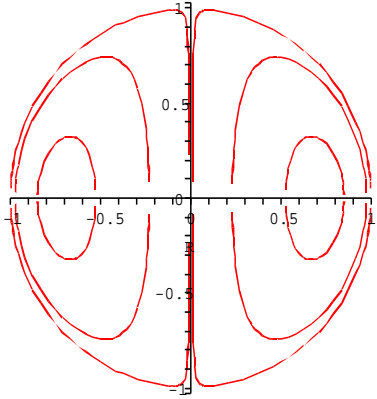}
\caption{A meridional cut of a spherical star with an assumed
electron density profile $n(r)=n_0[1-(r/r_0)^2]$, where $r$ is the
(spherical) radial coordinate, $r_0$ is the stellar radius, and
$n_0$ is the electron density at $r=0$. Shown are
``$\chi$-surfaces'' corresponding to $\chi/\chi_0=5$, 20, and
$10^4$, where $\chi_{0}=c/(4\pi e n_{0} r_{0}^{2})$.}
\end{figure}

In the absence of dissipation, the evolution of the magnetic field
at any given point depends only on its value at other points on
the same $\chi$-surface, not on the values on adjacent surfaces.
We define a new coordinate $s$ on each $\chi$-surface by the
condition $\partial/\partial s\equiv
R^{2}\nabla\chi\times\nabla\phi\cdot\nabla$, so a variation $ds$
at constant $\chi$ and $\phi$ corresponds to a physical
displacement $|d\mathbf{r}|=R|\nabla\chi|ds$. This allows the
field evolution to be written as the dissipationless Burgers
equation \citep{Burgers40,Burgers48},
\begin{equation}
\mathcal{B}_{t}+\mathcal{B}\mathcal{B}_{s}=0,
\end{equation}
with the subscripts denoting partial derivatives. The domain of this
equation is a closed loop of constant $\chi$ and $\phi$, therefore
its boundary conditions must be periodic.

It is well known \citep{Burgers40,Burgers48} that, for a given
initial condition $\mathcal{B}(s,0)=f(s)$, where $f$ is an
arbitrary function, Burgers' equation has an implicit, analytic
solution,
\begin{equation}\label{Burgers-solution}
\mathcal{B}=f(s-\mathcal{B}t),
\end{equation}
so the value $\mathcal{B}$ of the function at any point $s$ is
carried along the surface with ``velocity'' $ds/dt=\mathcal{B}$. The
larger values travel faster, so discontinuities are formed when they
are about to ``overtake'' the more slowly moving, smaller values.

The model considered here encompasses two previously studied
models as limiting cases. One is a plane-parallel slab with a
horizontal magnetic field and vertically decreasing electron
density, in which \citet{VCO} showed that Burgers' equation
governs the field evolution on horizontal planes. The other is a
sphere or other axisymmetric body with a toroidal field and {\it
uniform} electron density, in which simulations \citep{Urpin}
yield that the magnetic field (initially taken to be symmetric
with respect to the equatorial plane) progressively concentrates
in one hemisphere, where it develops a strong gradient and
eventually gets dissipated. This is easy to understand in terms of
our results, as in this case the $\chi$-surfaces are coaxial,
cylindrical surfaces, along which the field drifts towards the
surface, developing a discontinuity characteristic of Burgers'
equation \citep{Araya,Pons}.

One feature of our, more general solution is not present in either
of these special cases. Since the quantity $\mathcal{B}$ is
conserved as it is carried along the $\chi$-surfaces, the magnetic
field strength
$|\mathbf{B}|=|\mathcal{B}\nabla\phi|=|\mathcal{B}|/R$ changes in
inverse proportion to the distance to the axis, $R$. In
particular, when the flux near the surface of the star is carried
from low latitudes to the vicinity of a pole, the field strength
increases, as found in simulations
\citep{Hollerbach02,Hollerbach04}.

In both limiting cases mentioned above, one is forced to adopt
somewhat arbitrary boundary conditions, in the first case because
the slab is infinite, in the other because the electron density
drops abruptly to zero at the surface of the star, where the
$\chi$-surfaces end. This is not a problem in more realistic
cases, in which the electron density decays continuously to zero
at the stellar surface, and $\chi$-surfaces are closed and
therefore have no boundaries. Of course, a real neutron star is
not fully solid, but has a fluid core, which will be threaded by
the $\chi$-surfaces. Thus, magnetic flux will generally be
transported back and forth between the fluid core and the solid
crust, an effect that will require understanding the Hall drift in
fluid matter, which is addressed by \citet{RT07}.

In order to find the time of occurrence of the discontinuities in
the magnetic field, one may consider the partial derivative of eq.
(\ref{Burgers-solution}) with respect to $s$,
$\mathcal{B}_{s}=(1-\mathcal{B}_{s}t)f'(s-\mathcal{B}t)$, from
which one obtains $\mathcal{B}_{s}=f'/(1+tf')$. A discontinuity
($\mathcal{B}_{s}\rightarrow \pm \infty$) is formed at time
$t_{disc}=1/\max (-f')$, at the comoving point where the initial
spatial derivative had the largest, negative slope. (Note that the
Hall effect is not invariant under spatial reflection.)

These discontinuities in the magnetic field strength correspond to
sufaces of infinite current density. In practice, before a
``current sheet'' becomes singular, resistive dissipation must
occur, even for arbitrarily small $\eta$. This dissipation occurs
in a thin layer, whose width is $\approx
\eta/(R|\nabla\chi|\Delta\mathcal{B})$, where $\Delta\mathcal{B}$
is the difference in the values of $\mathcal{B}$ across the (near)
discontinuity \citep{VCO}. However, the rate of dissipation
depends exclusively on the rate at which the Hall effect
transports the scalar variable $\mathcal{B}$ (related to the
magnetic flux) to the current sheet.


\subsection{Conserved quantities} \label{sec:conserved}

From the induction equation for a toroidal field
(eq.~[\ref{toroidalinduction}]), and using the identity
\begin{equation}\label{identity}
\nabla\cdot\left(\mathbf{w}\over
R^2\mathcal{B}\right)=\nabla\cdot(\nabla\chi\times\nabla\phi)=0,
\end{equation}
we can show that any scalar function $F(\mathcal{B})$ defines a
``density'',
\begin{equation}\label{density}
\rho\equiv{F'\over R^2},
\end{equation}
and a ``flux'',
\begin{equation}\label{flux}
\mathbf{J}\equiv{\mathcal{B}\over R^2}{d\over
d\mathcal{B}}\left(F\over\mathcal{B}\right)\mathbf{w}-{\eta\over
R^2}F''\nabla\mathcal{B},
\end{equation}that satisfy a continuity equation
\begin{equation}\label{conservation}
{\partial\rho\over\partial t}+\nabla\cdot\mathbf{J}=-{\eta\over
R^2}F'''(\nabla\mathcal{B})^2,
\end{equation}
where primes ($'$) denote derivatives of $F$ with respect to its
argument, $\mathcal{B}$. Thus, the physical quantity whose density
is $\rho$ is transported by the Hall drift in the direction of
$\mathbf{w}$, i.~e., along $\chi$-surfaces, and by Ohmic diffusion
in the direction of decreasing $\mathcal{B}$, while it is being
destroyed by the right-hand-side term, which is also related to
Ohmic diffusion. (For definiteness, in this discussion we assumed
that $\mathcal{B}$, $F'$, $F''$, and $F'''$ are all $>0$.) Thus,
in the absence of Ohmic diffusion ($\eta=0$), the volume integral
over the toroid bounded by a $\chi$-surface,
\begin{equation}\label{Q}
Q(\chi_0;t)\equiv\int_{\chi(\mathbf{r})<\chi_0}\rho(\mathcal{B}[\mathbf{r},t])~dV,
\end{equation}
is constant in time. We consider two specific examples of such
conservation laws, namely those of magnetic flux and of magnetic
energy.

In order to obtain magnetic flux conservation, we must set
$F=\mathcal{B}^2/2$. For this choice, the right-hand-side of
equation~(\ref{conservation}) vanishes, indicating that magnetic
flux can be transported by both Hall drift and Ohmic diffusion,
but neither created nor destroyed, and the continuity equation can
be rewritten in terms of the magnetic flux density
$B=\mathcal{B}/R$ as
\begin{eqnarray}\label{fluxconservation}
{\partial B\over\partial t}&+&{\partial\over\partial
R}\left[{1\over 2}Bw_R-{\eta\over R}{\partial\over\partial
R}(RB)\right]\\
&+&{\partial\over\partial z}\left[{1\over 2}Bw_z-\eta{\partial
B\over\partial z}\right]=0,
\end{eqnarray}
which can be interpreted as a two-dimensional continuity equation
for $B$ on the Cartesian $R-z$ plane.

For magnetic energy conservation, we set
$F=\mathcal{B}^3/(24\pi)$, obtaining
\begin{eqnarray}\label{energyconservation}
{\partial\over\partial t}\left[{1\over 8\pi}\left(\mathcal{B}\over
R\right)^2\right]&+&\nabla\cdot\left[\left(\mathcal{B}\over
R\right)^2{\mathbf{w}\over 12\pi}-{\eta\mathcal{B}\over 4\pi
R^2}\nabla\mathcal{B}\right]\\&=&-{\eta\over 4\pi
R^2}(\nabla\mathcal{B})^2.
\end{eqnarray}
The right-hand-side is negative-definite, so magnetic energy can
be destroyed by Ohmic diffusion, but not created. We note that the
magnetic energy flux vector obtained here is generally different
(in its Hall-drift part) from the standard Poynting vector,
\begin{equation}\label{Poynting}
\mathbf{S}={c\over
4\pi}\mathbf{E}\times\mathbf{B}=-{\chi\mathcal{B}\over
4\pi}\nabla\mathcal{B}\times\nabla\phi-{\eta\mathcal{B}\over 4\pi
R^2}\nabla\mathcal{B}.
\end{equation}
However, their divergence, which determines their physical
effects, is the same.

In the scenario discussed in the previous section, in which Hall
drift dominates and Ohmic dissipation is only significant in thin
current sheets, we expect no large-scale transport of magnetic
flux or energy across $\chi$-surfaces, but only a redistribution
on each surface by Hall drift and local Ohmic dissipation in the
current sheets, which causes the magnetic energy to decrease while
keeping the magnetic flux constant. Thus, the asymptotically
resulting field configurations
$\mathbf{B}_{0}=\mathcal{B}(\chi)\nabla\phi$ are minima of the
magnetic energy subject to conservation of flux in the region
within each $\chi$-surface, and are the only toroidal equilibrium
states as far as the Hall effect is concerned. We note that, like
the stationary, poloidal field found by \citet{Cumming},
corresponding to a rigidly rotating electron fluid, these
configurations are also not force-free, in the sense that
$\mathbf{j}\times\mathbf{B}\neq 0$. In the long term, of course,
diffusion will allow magnetic flux transport across
$\chi$-surfaces, eventually leading to magnetic field decay.


\section{Perturbations and instability}
\label{sec:instability}


\subsection{Poloidal perturbations of the toroidal equilibrium
field} \label{sec:poloidal}

We now study the stability of this stationary, toroidal magnetic
field to a small, poloidal perturbation, $\mathbf{B}_{1}$. The
associated velocity field $\mathbf{v}_1$ is toroidal, therefore
$\mathbf{v}_{1}\times\mathbf{B}_{0}=0$, and the linearized
evolution equation for the perturbation reduces to
\begin{equation}
\frac{\partial\mathbf{B}_{1}}{\partial
t}=\nabla\times(\mathbf{v}_{0}\times\mathbf{B}_{1}),
\end{equation}
which implies that the field lines of the perturbation are carried
along by the background electron flow field,
$\mathbf{v}_{0}=-(c/4\pi ne)\nabla\times\mathbf{B}_0$, and the
perturbation field $\mathbf{B}_1$ remains poloidal.

Thus, it can be written as
$\mathbf{B}_{1}=\nabla\mathcal{A}(\chi,s,t)\times\nabla\phi$, with
$|\nabla\mathcal{A}|\ll|\mathcal{B}|$. Replacing into the above
equation, one finds that the potential evolves according to
\begin{equation}
{\partial\mathcal{A}\over\partial
t}=-\mathbf{v}_0\cdot\nabla\mathcal{A}=\chi{d\mathcal{B}\over
d\chi}{\partial\mathcal{A}\over\partial s},
\end{equation}
i.~e., the scalar quantity $\mathcal{A}$ is also carried along
$\chi$-surfaces by the unperturbed electron flow. Thus, it can
generally be written in terms of its initial condition as
\begin{equation}
\mathcal{A}(\chi,s,t)=\mathcal{A}(\chi,s+\chi[d\mathcal{B}/d\chi]t,0).
\end{equation}

Initially, the smallness of $\mathbf{B}_1$ forces $\mathcal{A}$ to
be a smooth function of position, taking similar values on
adjacent points of different $\chi$-surfaces. However, when the
circulation periods of the electrons around different
$\chi$-surfaces are different, each point comoving with them will
come close to others that were initially far away, so the values
of $\mathcal{A}$ within a vicinity of fixed size around a given
point will become progressively different, leading to a linearly
increasing perturbation field. Another way of viewing this process
is by realizing that the magnetic field lines of the perturbation
are stretched as different parts are carried by electron currents
circulating with different periods on different $\chi$-surfaces. A
simulation of this evolution is shown in Fig.~2.

\begin{figure}
  \centering
    \includegraphics[width=4.2cm,height=8.4cm]{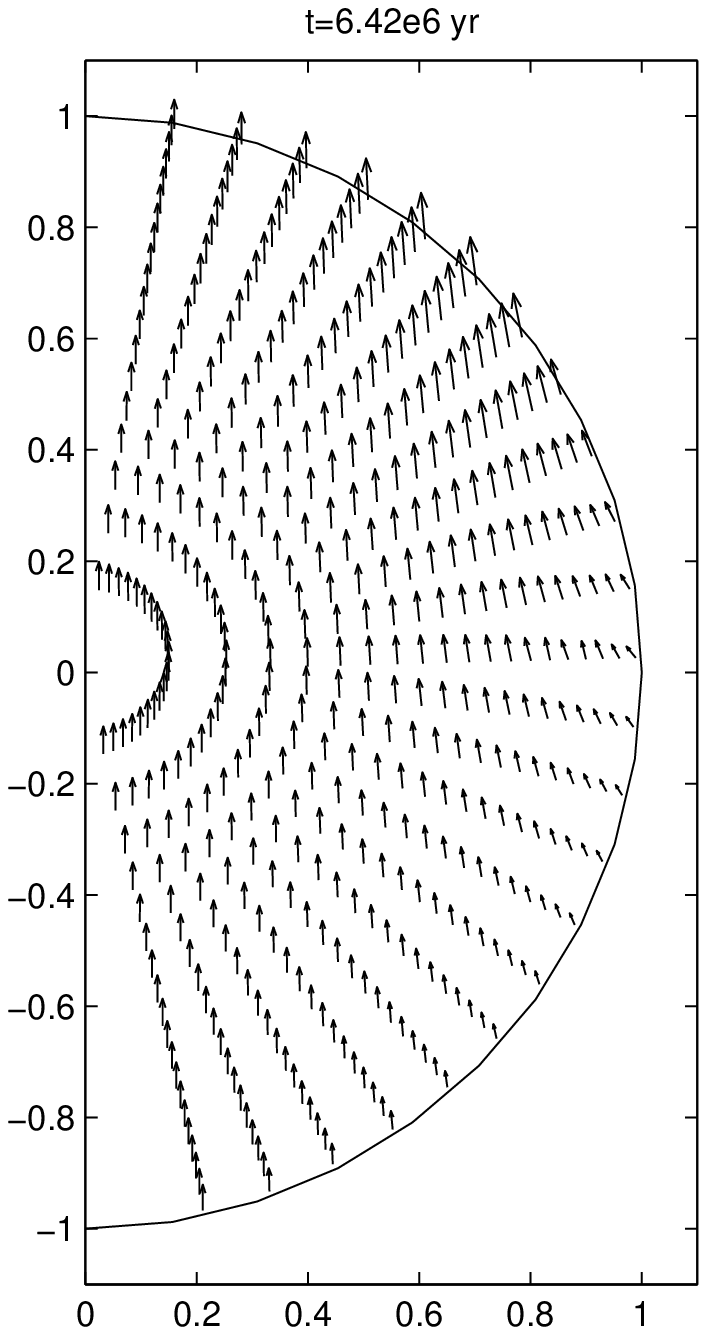}
    \includegraphics[width=4.2cm,height=8.4cm]{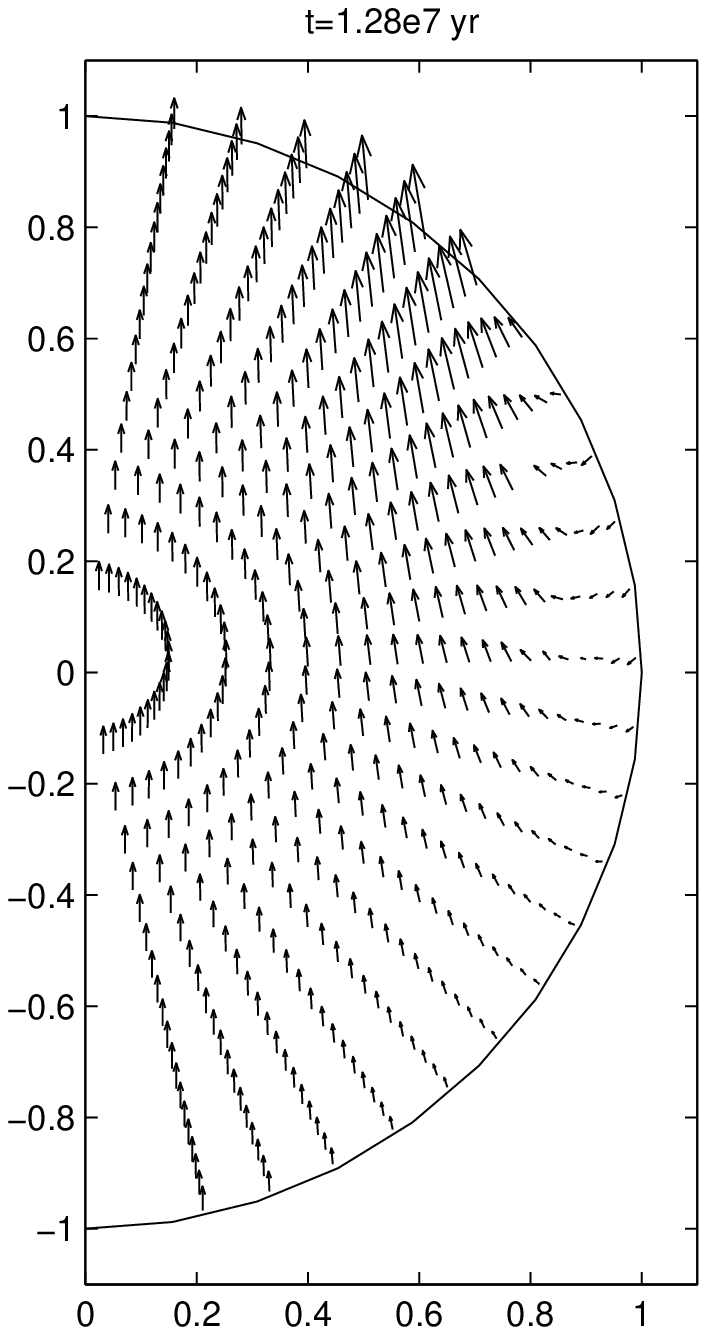}\\
    \includegraphics[width=4.5cm,height=8.4cm]{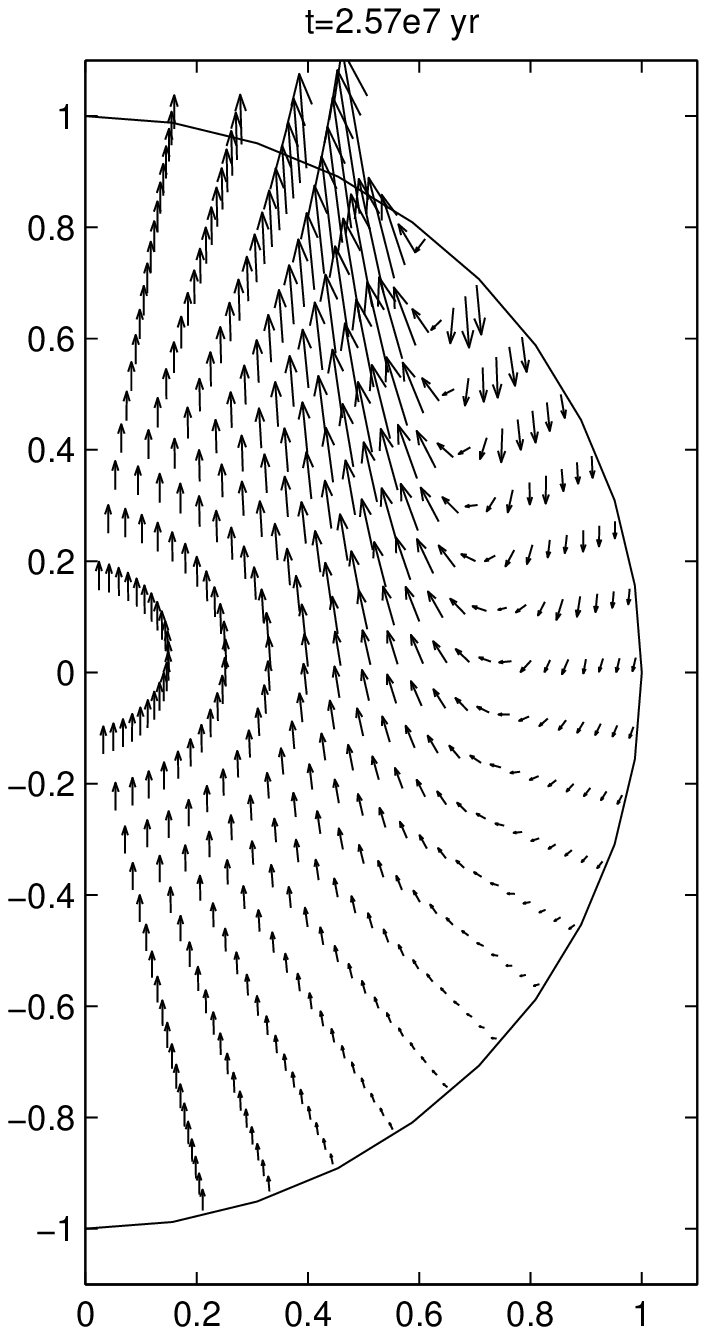}
    \includegraphics[width=4.2cm,height=8.4cm]{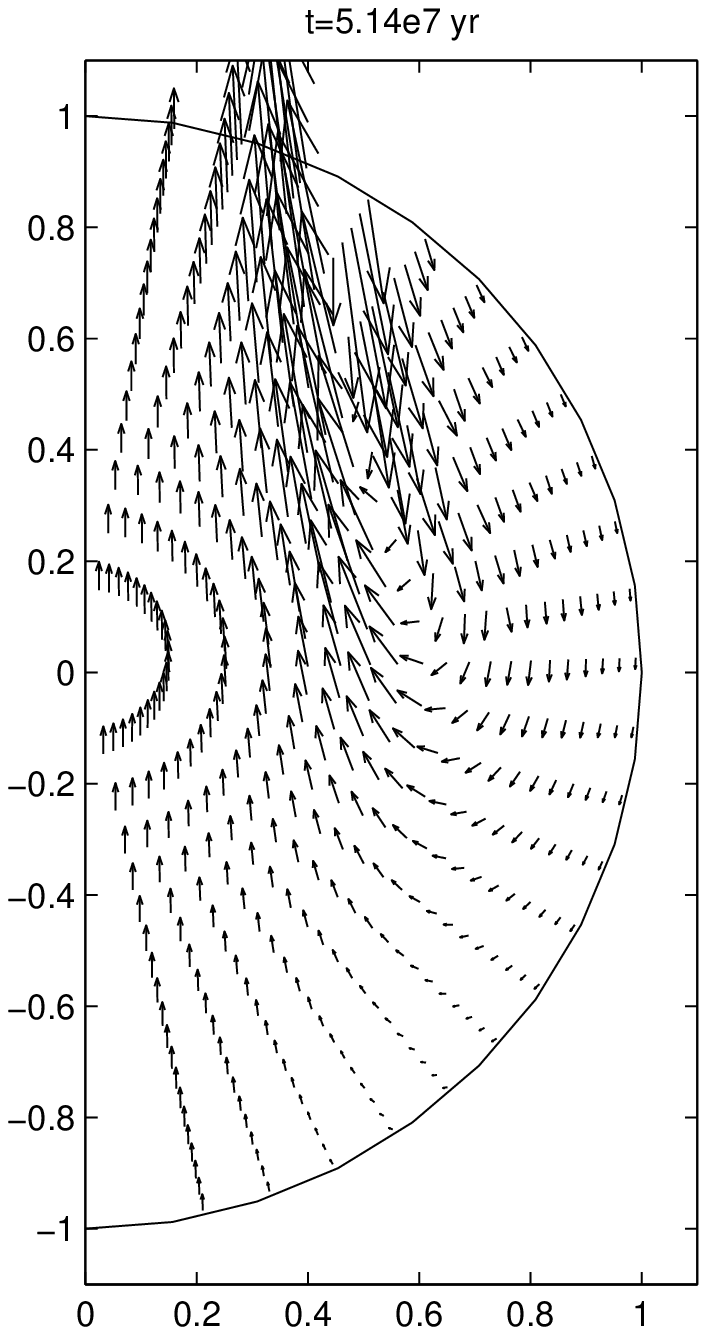}
\caption{The evolution of a weak poloidal magnetic field component
affected by a much stronger toroidal background field. Shown are
the poloidal field vectors on a cut through a star with electron
density profile $n(r)=n_0[1-(r/r_0)^2]$, with central density
$n_{0}=10^{36}\mathrm{cm^{-3}}$, and stellar radius
$r_{0}=10\mathrm{km}$. The background toroidal field has the form
$\mathbf{B}_0=\mathcal{B}(\chi)\nabla\phi$, where we chose
$\mathcal{B}(\chi)=b_{0}r_{0}(\chi_{0}/\chi)^2$, with
$b_0=10^{14}\mathrm{G}$, $\chi=c/(4\pi neR^2)$, and $\chi_0$ its
minimum value. The initial poloidal field was chosen uniform and
pointing along the symmetry axis. The evolutionary times scale
$\propto n_0 r_0^2/B_0$.}
\end{figure}

As the function $\mathcal{A}(\chi,s,t)$ is periodic, the perturbed
field oscillates in time, periodically for the component
perpendicular to the $\chi$-surfaces, and with increasing
amplitude for the tangential component. A similar, oscillating
behavior has been observed in numerical simulations
\citep{Hollerbach02,Hollerbach04,Shalybkov}. However, given their
more complex geometries, a direct connection is difficult to
establish.

\subsection{Plane-parallel analog} \label{sec:plane}

In order to understand the relation of the instability found
analytically by us to those found numerically by
\citet{Rheinhardt02} in a plane-parallel slab and later discussed
by \citet{Cumming}, we consider the plane-parallel geometry
studied by these authors, in which a horizontal field varying
nonlinearly with depth generates different velocities at different
depths, stretching a small, vertical field component.

We take the vertical coordinate to be $z$, and consider a
background magnetic field ${\bf B}_0=f(z)\hat x$, which generates
an electron velocity ${\bf v}_0=-cf'(z)/[4\pi n(z)e]\hat y$, where
primes denote derivatives with respect to $z$. If an additional,
uniform, vertical field $B_{1z}$ is introduced, it causes the
linear growth of a component in the $y$ direction, $\partial
B_{1y}/\partial t=-[cB_{1z}f'(z)/(4\pi n(z)e)]'$. In some sense,
this is a particular version or limiting case of the instabilities
found by \citet{Rheinhardt02}, with vanishing horizontal wave
vector. However, it does not appear in their calculation, as this
field perturbation cannot be written in terms of two sinusoidally
varying potentials, as done in their equations (4) and (5). For
that representation, a magnetic field perturbation with vanishing
horizontal wave vector would be identically zero. Nevertheless, in
our solution as in that work, if the electron density is taken to
be constant, the growth of the perturbation depends directly on
the second spatial derivative of the background field strength,
$f''(z)$, suggesting that we are in fact seeing the same kind of
instability. In our model, we confirm the conclusion of
\citet{Cumming} that this finite second derivative is caused by a
sheared electron flow velocity, which bends the vertical field
component, creating an additional horizontal field.

Curiously, in this simple geometry, we can actually calculate the
nonlinear development of the instability. We consider a field with
three components of arbitrary strength: $B_x(z,t)$, which is
initially the ``background'' field, but is now allowed to evolve
arbitrarily large perturbations; the initially small (or
vanishing), but growing component $B_y(z,t)$; and the vertical
component $B_z$, which represents the initial perturbation. For
this choice, the components of the induction equation become
\begin{eqnarray}\label{whistler_eq}
{\partial B_x\over\partial t}&=&{\partial\over\partial
z}\left({cB_z\over 4\pi n(z)e}{\partial B_y\over\partial
z}\right),\\
{\partial B_y\over\partial t}&=&-{\partial\over\partial
z}\left({cB_z\over 4\pi n(z)e}{\partial B_x\over\partial
z}\right),\\
{\partial B_z\over\partial t}&=&0.
\end{eqnarray}
These are the equations for ``helicons'' or ``whistler waves'' of
arbitrary amplitude propagating along the $z$ axis. For
$n(z)=\mathrm{constant}$, they have a dispersion relation
$\omega=[cB_z/(4\pi ne)]k^2$, where $\omega$ is the frequency and
$k$ is the magnitude of the wave vector (which points in the $\pm
z$ direction). One particular solution is
\begin{eqnarray}\label{whistler_sol}
B_x(z,t)&=&B_m\cos(kz)\cos(\omega t),\\
B_y(z,t)&=&B_m\cos(kz)\sin(\omega t),\\
B_z&=&\mathrm{constant},
\end{eqnarray} where the horizontal component initially points along
the $x$ axis, then turns to the $y$ axis at a rate determined by
$B_z$, and eventually turns around completely. Thus, the evolution
of the field in this case is strictly periodic (with the period
equal to the Hall time scale of the ``small'' component $B_z$) and
does not lead to dissipation on a Hall time scale. If Ohmic
diffusion is introduced, the amplitude of the horizontal component
will decay exponentially, on the resistive time scale $1/(\eta
k^2)$. However, we suspect that this ``clean'' behavior is a
peculiarity of this very symmetric configuration, and will not
hold in more realistic cases, including the simulations of
\citet{Rheinhardt02}.


\subsection{Energy conservation in Hall instabilities} \label{sec:energy}

In this section, we would like to clarify an issue that was
raised, but in our view not fully clarified, by both
\citet{Rheinhardt02} and \citet{Cumming}, namely the conservation
of energy in the growth of Hall instabilities.

In any slow evolution of a magnetic field, one can use the
induction equation and Amp\'ere's law to show that the magnetic
energy density $u=B^2/(8\pi)$ satisfies
\begin{equation}\label{continuity}
{\partial u\over\partial t}+\nabla\cdot{\bf S}=-{\bf j}\cdot{\bf
E},
\end{equation}
with the Poynting flux vector ${\bf S}=(c/4\pi){\bf E}\times{\bf
B}$. For pure Hall drift, the electric field, ${\bf E}={\bf
j}\times{\bf B}/(ne)$, is perpendicular to the current, so the
right-hand side vanishes and the magnetic energy satisfies an
exact conservation law \citep{GR92}.

If we decompose the magnetic field, ${\bf B}={\bf B}_0+{\bf B}_1$,
where ${\bf B}_0$ is the constant or slowly varying ``background
field'', whereas ${\bf B}_1$ is a small, more rapidly varying
``perturbation'', each can be taken to satisfy an (exact)
induction equation, $\partial{\bf B}_\alpha/\partial
t=-c\nabla\times{\bf E}_\alpha$, where
\begin{eqnarray}\label{electric}
{\bf E}_0&=&{1\over nec}{\bf j}_0\times{\bf B}_0,\\
{\bf E}_1&=&{1\over nec}\left({\bf j}_0\times{\bf B}_1+{\bf
j}_1\times{\bf B}_0+{\bf j}_1\times{\bf B}_1\right),
\end{eqnarray}
and ${\bf j}_\alpha=(c/4\pi)\nabla\times{\bf B}_\alpha$, for
$\alpha=0,1$. The magnetic energy density can be decomposed into
three parts, $u=u_0+u_1+u_2$, where $u_0=B_0^2/(8\pi)$ and
$u_2=B_1^2/(8\pi)$ are positive-definite, while $u_1={{\bf
B}_0\cdot{\bf B}_1}/(4\pi)$ is not. For $u_0$, the derivation
above can be retraced to show that it satisfies an exact
conservation law. However, the evolution of $u_1$ and $u_2$ yields
\begin{eqnarray}
\label{u1}
{\partial u_1\over\partial t}+\nabla\cdot{\bf S}_1&=&-{1\over
nec}{\bf
j}_0\cdot{\bf j}_1\times{\bf B}_1,\\
\label{u2}
{\partial u_2\over\partial t}+\nabla\cdot{\bf
S}_2&=&{1\over nec}{\bf j}_0\cdot{\bf j}_1\times{\bf B}_1,
\end{eqnarray}
where ${\bf S}_1=(c/4\pi)({\bf E}_1\times{\bf B}_0+{\bf
E}_0\times{\bf B}_1)$ and ${\bf S}_2=(c/4\pi){\bf E}_1\times{\bf
B}_1$. Thus, the volume integrals of the two pieces, $\int u_1$
and $\int u_2$, are not individually conserved, but their sum,
$\int (u_1+u_2)$, is conserved. To linear order in ${\bf B}_1$,
only $\int u_1$ is non-zero, and it is conserved. To quadratic
order, $\int u_2$ can increase (in the case of an instability),
being balanced by a corresponding decrease in $\int u_1$, which
becomes progressively more negative. Thus, instabilities can only
occur if, for some choices of the perturbation field
$\mathbf{B}_1$, the spatial integral of the right-hand side of
equation~(\ref{u2}) is positive. Clearly, this will not happen for
current-free field configurations ($\mathbf{j}_0=0$), for which
this right-hand side vanishes identically and no growth will
occur. Somewhat more generally, \citet{Cumming} have shown that
for their stationary configurations, in which the current density
is such that all the electrons rotate as a rigid body, the
volume-integral of the right-hand side is zero for all
perturbations that vanish on the boundaries, therefore also
proving their stability.


\subsection{Current-free, poloidal background field} \label{sec:current-free}

An application of the result of \S~\ref{sec:energy} is given by
the following, plausible scenario for the evolution of the
magnetic field in a neutron star. The field initially has an
approximately axisymmetric configuration with both poloidal and
toroidal components that are supported by currents that flow
partly in the solid crust and partly in the fluid core of the
star. In the core, the conductivity is very high, the Hall drift
is ineffective \citep{RT07}, and ambipolar diffusion not fast
enough, so the field is essentially frozen in some MHD-stable
configuration \citep{BS06}. In the solid crust, Ohmic dissipation
is more effective, particularly if aided by Hall drift, and one
might expect the currents to decay, so the field might end up in a
current-free configuration, which, by the arguments of the
previous section, is stable under the Hall drift.

In order to examine the evolution of the small perturbations, we
use our earlier notation, with the background field written as
$\mathbf{B}_0=\nabla\mathcal{A}_0\times\nabla\phi$, with
$\mathcal{B}_0=0$ (no toroidal component) and
$\Delta^*\mathcal{A}_0=0$ (no current). It clearly does not evolve
through Hall drift, since $\partial\mathcal{A}_0/\partial t=0$.
Adding a small perturbation, the latter will evolve as
\begin{eqnarray}
\label{dB1dt}{\partial\mathcal{B}_1\over\partial
t}=&R^2\mathbf{B}_0\cdot\nabla(\chi\Delta^*\mathcal{A}_1)&={1\over\chi}{\partial\over\partial\zeta}(\chi\Delta^*\mathcal{A}_1),\\
\label{dA1dt}{\partial\mathcal{A}_1\over\partial t}=&-\chi
R^2\mathbf{B}_0\cdot\nabla\mathcal{B}_1&=-{\partial\mathcal{B}_1\over\partial\zeta},
\end{eqnarray}
where we have defined the coordinate $\zeta$ along field lines of
$\mathbf{B}_0$ by the condition $\partial/\partial\zeta\equiv\chi
R^2\mathbf{B}_0\cdot\nabla=(c\mathbf{B}_0/4\pi ne)\cdot\nabla$,
analogous to the definition of $s$ on $\chi$-surfaces in
\S~\ref{sec:toroidal}.

One can combine the two equations, eliminating $\mathcal{B}_1$ and
obtaining
\begin{equation}
\label{d2A1dt} {\partial^2\mathcal{A}_1\over\partial
t^2}=-{\partial\over\partial
\zeta}\left[{1\over\chi}{\partial\over\partial
\zeta}(\chi\Delta^*\mathcal{A}_1)\right].
\end{equation}
This appears to be an hyperbolic differential equation with
wavelike solutions corresponding to helicons or whistler waves
travelling along the field lines of $\mathbf{B}_0$. In the WKB
approximation (small-wavelength perturbations), $\mathbf{B}_0$ can
be considered as uniform, $\Delta^*$ reduces to the Laplacian, and
this equation is exactly the whistler wave equation in a uniform
background field. The behavior of long-wavelength perturbations
may be more complex, but the arguments given above ensure that
they will also be stable.

The more general, ``uniformly rotating'' configuration of
\citet{Cumming} has $\mathbf{v}_0=\Omegav\times\mathbf{r}$, and
its field is therefore still poloidal, but not current-free. Since
$\chi\Delta^*\mathcal{A}_0=\Omega=\mathrm{constant}$ (although
non-zero), its perturbations satisfy exactly the same equations
(\ref{dB1dt}) through (\ref{d2A1dt}). This is consistent with the
energetic argument for it being stable as well.

This means that the current-free poloidal field (and its
generalization, the ``uniformly rotating'' field of
\citealt{Cumming}) is indeed stable under the Hall drift. The
field in a neutron-star crust might be able to settle into such a
state, which would remain in this form as long as the currents
supporting it in the core do not change.


\section{Variational principle}
\label{sec:variational}

Since the Hall drift conserves magnetic energy, it is tempting to
search for a variational principle that would yield Hall
equilibria, and in fact a fairly simple and natural one exists.
Consider stationary points of the magnetic energy $U=\int
B^2/(8\pi)$ subject to magnetic field perturbations
$\delta\mathbf{B}=\nabla\times(\xiv\times\mathbf{B})$ that are due
to an infinitesimal displacement field $\xiv$ that does not change
the electron density, i.~e., $\nabla\cdot(n\xiv)=0$. The latter
condition implies that the displacement field can be written as
$\xiv=(1/n)\nabla\times\mathbf{a}$, where $\mathbf{a}$ is an
arbitrary vector field. This allows a perturbation of the magnetic
energy density to be written as
\begin{eqnarray}\label{denergy}
\delta u&=&{\mathbf{B}\cdot\delta\mathbf{B}\over 4\pi}\\
&=&\nabla\cdot\left[{1\over
4\pi}(\xiv\times\mathbf{B})\times\mathbf{B}+{1\over
cn}(\mathbf{j}\times\mathbf{B})\times\mathbf{a}\right]\\
&&-\mathbf{a}\cdot\nabla\times\left(\mathbf{j}\times\mathbf{B}\over
cn\right).
\end{eqnarray}
Integrating over the volume, the divergence term becomes a surface
integral that can be made vanish by requiring that the normal
components of $\mathbf{j}$, $\mathbf{B}$ and $\xiv$ all vanish on
the surface. (The former two are required in order to ensure that
the energy in the volume is conserved under Hall drift, the latter
for consistency with the condition of not changing the electron
density.) With these conditions, in order to have $\delta U=0$ for
an otherwise arbitrary vector field $\mathbf{a}$, we must have
$\nabla\times(\mathbf{j}\times\mathbf{B}/n)=0$, which is also the
condition to have no Hall drift of the magnetic field. (It can be
shown that, for these perturbations and the adopted boundary
conditions, the magnetic helicity $\int\mathbf{A}\cdot\mathbf{B}$
is automatically conserved, i.~e., our constraint is stronger than
that of requiring helicity conservation.)

Put in a different way, our result states that, if the magnetic
field is such that it is not changed by the Hall drift,
$\xiv\propto(1/n)\nabla\times\mathbf{B}$, then its energy will not
change under the more general class of displacements
$\xiv=(1/n)\nabla\times\mathbf{a}$, with arbitrary $\mathbf{a}$.

Presumably, a subset of the configurations at which the magnetic
energy is stationary, subject to the constraints discussed, will
be maxima or minima of the magnetic energy, i.~e., to order
$\xiv^2$ or higher, all perturbations will cause magnetic energy
changes of the same sign. For such configurations, there will be
only a small set of nearby configurations with a similar energy,
and thus they will be stable under Hall drift. The other
configurations at which the magnetic energy is stationary will
correspond to saddle points, inflection points, etc., whose energy
is the same as that of a large set of other configurations that
can be obtained by continuously deforming the magnetic field
structure. We expect that these will generally be Hall-unstable,
in the sense that a small, initial perturbation can eventually
make them evolve into a very different configuration.


\section{Conclusions}
\label{sec:conclusions}

We have found some interesting results regarding the Hall-drift
evolution of an axially symmetric magnetic field in solid matter,
in which electrons are the only mobile particles:

1) We have characterized ``Hall equilibrium'' configurations that
do not evolve under the Hall drift (\S~\ref{sec:axial}).

2) We have found that an exactly toroidal field evolves
discontinuities described by the Burgers equation, which dissipate
on the Hall drift timescale (\S~\ref{sec:Burgers}), as in the
plane-parallel case studied earlier by \citet{VCO}. The field
evolves into a stationary state that minimizes magnetic energy
subject to flux conservation within a set of nested, toroidal
``$\chi$-surfaces'' (\S~\ref{sec:conserved}).

3) However, we found these stationary, toroidal fields to be
unstable to poloidal perturbations (\S~\ref{sec:poloidal}), making
them an unrealistic model for the fields of neutron stars. The
instability involved is due to shearing of the perturbation field
by the background electron velocity, and thus closely related to
those studied by \citet{Rheinhardt02} and \citet{Cumming}.

4) We have discussed the issue of energy conservation in Hall
instabilities, giving a criterion for their occurrence and showing
that current-free fields are stable (\S~\ref{sec:energy}). We
applied it to argue that a current-free, poloidal field in the
crust, supported by axially symmetric currents flowing only in the
core of the star, might represent a long-lived magnetic
configuration of a neutron star (\S~\ref{sec:current-free}).

5) Finally, we showed that a Hall equilibrium is a stationary
point for the magnetic energy, subject to displacements that do
not alter the electron density, and used this to argue that maxima
or minima of the magnetic energy with respect to such
perturbations will be stable under the Hall drift
(\S~\ref{sec:variational}).

\acknowledgements{This work was supported by FONDECYT Regular
Grants 1020840 (A.~R. and J.~P.~P.), 1020844 (R.~B. and P.~A.~A.),
1060644 (A.~R.), and 1060651 (R.~B.). J.~P.~P. is also supported
by a doctoral fellowship at Pontificia Universidad Cat\'olica de
Chile (PUC). A visit by D.~L. to PUC was made possible by FONDECYT
International Cooperation Grant 7020840. The authors thank R.
Fern\'andez, U. Geppert, M. Lyutikov, H. Spruit, and C. Thompson
for useful discussions.}

\end{document}